\documentclass{article}

\usepackage{arxiv}

\usepackage[utf8]{inputenc} 
\usepackage[T1]{fontenc}    
\usepackage{hyperref}       
\usepackage{url}            
\usepackage{booktabs}       
\usepackage{amsfonts}       
\usepackage{nicefrac}       
\usepackage{microtype}      
\usepackage{cleveref}       
\usepackage{graphicx}
\usepackage{cite}
\usepackage{doi}
\usepackage{multirow}
\usepackage[dvipsnames,table]{xcolor}
\usepackage{bbding}
\usepackage{authblk}
\usepackage{longtable}
\usepackage{setspace}
\usepackage{tikz}
\usepackage{enumitem}

\hypersetup{
    colorlinks,
    linkcolor={blue!90!black},
    citecolor={blue!90!black},
    urlcolor={blue!90!black}
}

\newcommand{\ballnumber}[1]{\tikz[baseline=(myanchor.base)] \node[circle,fill=.,inner sep=1pt] (myanchor) {\color{-.}\bfseries\footnotesize #1};}

\title{EDA-Aware RTL Generation\\with Large Language Models}

\date{}

\newbox{\orcid}\sbox{\orcid}{\includegraphics[scale=0.06]{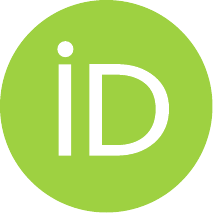}}

\author[1]{%
    Mubashir ul Islam%
}
\author[1]{%
    Humza Sami
}
\author[1,2]{%
    Pierre-Emmanuel Gaillardon
}
\author[1]{%
    Valerio Tenace\thanks{Corresponding author: \texttt{valerio@primis.ai}}
}
\affil[1]{PrimisAI, Los Gatos, CA, USA}
\affil[2]{University of Utah, Salt Lake City, UT, USA}

\renewcommand{\headeright}{}
\renewcommand{\undertitle}{}
\renewcommand{\shorttitle}{EDA-Aware RTL Generation with LLMs}

\hypersetup{
pdftitle={EDA-Aware RTL Generation with Large Language Models},
pdfsubject={cs.AI, cs.MA},
pdfauthor={Mubashir ul Islam, Humza Sami, Pierre-Emmanuel Gaillardon, Valerio Tenace},
pdfkeywords={Large Language Models, RTL Generation, Verification, Multi-Agent Systems, Generative AI, Electronic Design Automation},
}

\begin{document}
\maketitle

\setcounter{footnote}{0}

\begin{spacing}{1}
\begin{abstract}
    \textit{Large Language Models} (LLMs) have become increasingly popular for generating RTL code. However, producing error-free RTL code in a zero-shot setting remains highly challenging for even state-of-the-art LLMs, often leading to issues that require manual, iterative refinement. This additional debugging process can dramatically increase the verification workload, underscoring the need for robust, automated correction mechanisms to ensure code correctness from the start.

    In this work, we introduce {\scshape AIvril}2, a self-verifying, LLM-agnostic agentic framework aimed at enhancing RTL code generation through iterative corrections of both syntax and functional errors. Our approach leverages a collaborative multi-agent system that incorporates feedback from error logs generated by EDA tools to automatically identify and resolve design flaws. Experimental results, conducted on the VerilogEval-Human benchmark suite, demonstrate that our framework significantly improves code quality, achieving nearly a 3.4$\times$ enhancement over prior methods. In the best-case scenario, functional pass rates of 77\% for Verilog and 66\% for VHDL were obtained, thus substantially improving the reliability of LLM-driven RTL code generation.
\end{abstract}


\section{Introduction}
The rapid development of {\em Artificial Intelligence} (AI) has led to transformative changes across various industries, with {\em Large Language Models} (LLMs) standing out as powerful tools capable of generating human-like text and interpreting complex user instructions. Within the field of hardware design, LLMs hold the potential to revolutionize the entire design process, with projections suggesting that both front-end and back-end tasks could soon become fully automated~\cite{liu2023chipnemo,wu2024chateda}. Among the tasks gaining significant attention is the automated generation of {\em Register Transfer Level} (RTL) code. By interpreting user intent with minimal human input, LLMs can streamline workflows, effectively bridging the gap between conceptual design and physical implementation, thereby ultimately enhancing productivity.

However, despite their potential, the probabilistic nature of LLMs poses critical challenges, particularly in {\em zero-shot prompting} scenarios. In fact, without task-specific training, LLM outputs often contain syntactical and functional errors. While progress has been made in {\em Generative AI} (GenAI) for RTL design, concerns about the accuracy and reliability of generated code still remain. Frequently, the output requires substantial manual correction~\cite{tsai2023rtlfixer}, diminishing the efficiency promised by this new technology which instead often leads to an increased verification burden. This manual iterative correction process is time-consuming and can introduce additional errors, thus highlighting a critical gap in current solutions: the lack of robust, automated verification mechanisms enclosed within GenAI solutions.

In response, recent efforts have explored integrating multi-agent systems into RTL code generation workflows. These systems utilize collaborative agents that leverage feedback from {\em Electronic Design Automation} (EDA) tools to iteratively debug both syntactical and functional errors~\cite{tsai2023rtlfixer,huang2024towards}. However, most current solutions are limited, addressing isolated issues, e.g., such as syntax correction or partial debugging, without offering a fully integrated LLM-driven code generation pipeline. Additionally, many approaches are optimized for a specific RTL language, usually Verilog~\cite{liu2023chipnemo,liu2023verilogeval,zhao2024codev}, which restricts their applicability to different hardware description languages.

In this work, we build upon our previous work~\cite{sami2024aivril} to introduce {\scshape AIvril}2, a self-verifying, LLM-agnostic agentic framework designed to enhance LLM-driven RTL code generation by iteratively correcting both syntactical and functional errors. The key contributions of this paper are threefold:
\setlist{nolistsep}
\begin{itemize}[noitemsep]
    \itemsep0em
    \item We present a new two-stage testbench and RTL code generation pipeline, with the first stage focused on syntactical corrections and the second on functional corrections. Both stages, or optimization loops, integrate LLMs with EDA tools, enabling continuous feedback between agents and tools for progressive code refinement.
    \item We detail the roles and behaviors of each specialized LLM-based agent: the {\em Code Agent}, responsible for generating robust RTL code and comprehensive testbenches; the {\em Review Agent}, which interprets complex EDA logs to detect and correct syntactical errors; and the {\em Verification Agent}, tasked with analyzing functional traces to ensure design accuracy.
    \item We demonstrate that {\scshape AIvril}2 is fully orthogonal to the target RTL language and completely LLM-agnostic, making it adaptable to various scenarios, thus enhancing its versatility for different and diverse workflows.
\end{itemize}

Experimental results, conducted on the VerilogEval-Human benchmark suite, demonstrate that {\scshape AIvril}2 achieves a 3.4$\times$ improvement over existing solutions in the best case, with pass@1 rates of 77\% for Verilog and 66\% for VHDL, underscoring the effectiveness of our proposed language-agnostic design.

The remainder of this paper is organized as follows: Section~\ref{sec:background} provides background information and discusses related work, highlighting the limitations of existing approaches. Section~\ref{sec:methods} details the internal mechanisms of {\scshape AIvril}2, including its multi-agent architecture and self-verification mechanisms. Section~\ref{sec:results} presents the experimental results, showcasing the advantages of our solutions over existing LLM-driven RTL generation techniques. Finally, Section~\ref{sec:conclusions} concludes the paper with a summary of the findings, suggesting potential directions for future research in enhancing automated RTL code verification.

\section{Background \& Related Work}\label{sec:background}
Decision-making frameworks for multi-agent systems are set to substantially influence GenAI-driven methodologies in hardware design. This section provides an overview of how verbal reasoning and action planning interact in autonomous systems, underscoring their growing role in GenAI applications for RTL design. Recent breakthroughs and key challenges in this evolving field are also highlighted.

\subsection{Multi-Agent Systems}
The integration of verbal reasoning with decision-making processes in autonomous systems has been a significant focus of recent research. LLMs have demonstrated their ability to manage multi-step reasoning, solving tasks such as arithmetic, commonsense reasoning, and symbolic operations~\cite{wei2022chain}. By decomposing complex problems into sequential steps---an approach commonly referred to as chain-of-thought prompting---LLMs effectively enhance the reasoning process. However, a notable limitation of these models is their dependence on internal reasoning without factoring-in any external data. As the reasoning chain lengthens, the risk of errors or hallucinations increases, which can compromise the output reliability. Without external validation, models may generate responses that, while seemingly plausible, are ultimately inaccurate.

To mitigate these shortcomings, recent efforts have explored LLM agents in interactive environments where predictions and action planning are based on real-time observations~\cite{huang2022language}. A notable paradigm in this domain is ReAct~\cite{yao2022react}, which couples reasoning with subsequent actions. This combination enables the model to plan and monitor actions while adapting dynamically to inputs received by the environment it operates in. This reasoning process supports the model in refining its plans, while external actions allow for interaction with real-world sources, such as databases, sensors, or other agents. This real-world feedback helps validate internal reasoning, reducing errors and hallucinations. As discussed in the next section, GenAI solutions for RTL design have increasingly adopted similar approaches, incorporating reasoning and action planning into unified frameworks. As a result, these strategies have enhanced the reliability and quality of RTL code, addressing key challenges in automated hardware design.

\subsection{GenAI for RTL Design}\label{sec:related-work}
Chip-Chat~\cite{blocklove2023chip} probably represents the first pioneering attempt at reproducing a fully automated hardware design workflow, from initial design to tapeout, by employing general-purpose LLMs like ChatGPT throughout the entire process. Since then, with the advent of advanced LLMs, zero-shot RTL code generation has significantly improved. These modern models leverage their language capabilities to effectively convert specifications into RTL code, though their performance still falls short compared to what they achieve with other programming languages. Therefore, given the stringent reliability requirements in RTL design, recent efforts have increasingly focused on enhancing the quality-of-results of LLM-driven solutions.

Prior work targets specific phases of RTL generation, such as syntax correction, partial debugging, or specific error handling. Notable approaches include domain-adapted LLMs (e.g., ChipNeMo~\cite{liu2023chipnemo}), data augmentation techniques~\cite{chang2024data}, and Retrieval-Augmented Generation (RAG)\cite{tsai2023rtlfixer}. While these methods provide notable improvements, they often lack a unified framework to integrate these advancements into a cohesive LLM-driven pipeline. For instance, RTLFixer\cite{tsai2023rtlfixer} introduced a ReAct and RAG-based approach that leverages error logs to iteratively correct syntax errors. Other examples include VeriAssist~\cite{huang2024towards} and VerilogCoder~\cite{ho2024verilogcoder}, which interactively interface LLMs with simulation tools to enable self-correction and self-verification mechanisms. VeriAssist, however, shows degraded performance when it only relies on self-generated testbenches. On the other hand, VerilogCoder does not generate testbenches as part of the verification process, therefore limiting its applicability to specific cases where external validation is already available. More recently, we introduced {\scshape AIvril}~\cite{sami2024aivril}, aiming to provide a more robust framework for autonomously generating and verifying RTL code alongside testbenches. However, the simultaneous generation of RTL and testbenches introduces additional complexity into the overall process.

Building upon~\cite{sami2024aivril}, the proposed {\scshape AIvril}2 framework adopts a testbench-first methodology, where an exhaustive testbench is first self-generated and then leveraged throughout the optimization phases. Moreover, our tool is designed to be language-agnostic, offering versatility across different RTL languages. By incorporating a multi-agent, ReAct-based mechanism, we seamlessly integrate verification processes into the LLM-driven design flow, significantly improving reliability and flexibility. This comprehensive approach addresses current limitations by providing a holistic and adaptable framework for RTL code generation, ultimately enhancing the quality and applicability of LLM-generated designs across various hardware contexts.

\section{{\scshape AIvril}2: EDA-Aware RTL Design}\label{sec:methods}

\begin{figure*}[!ht]
    \centering
    \includegraphics[width=\textwidth]{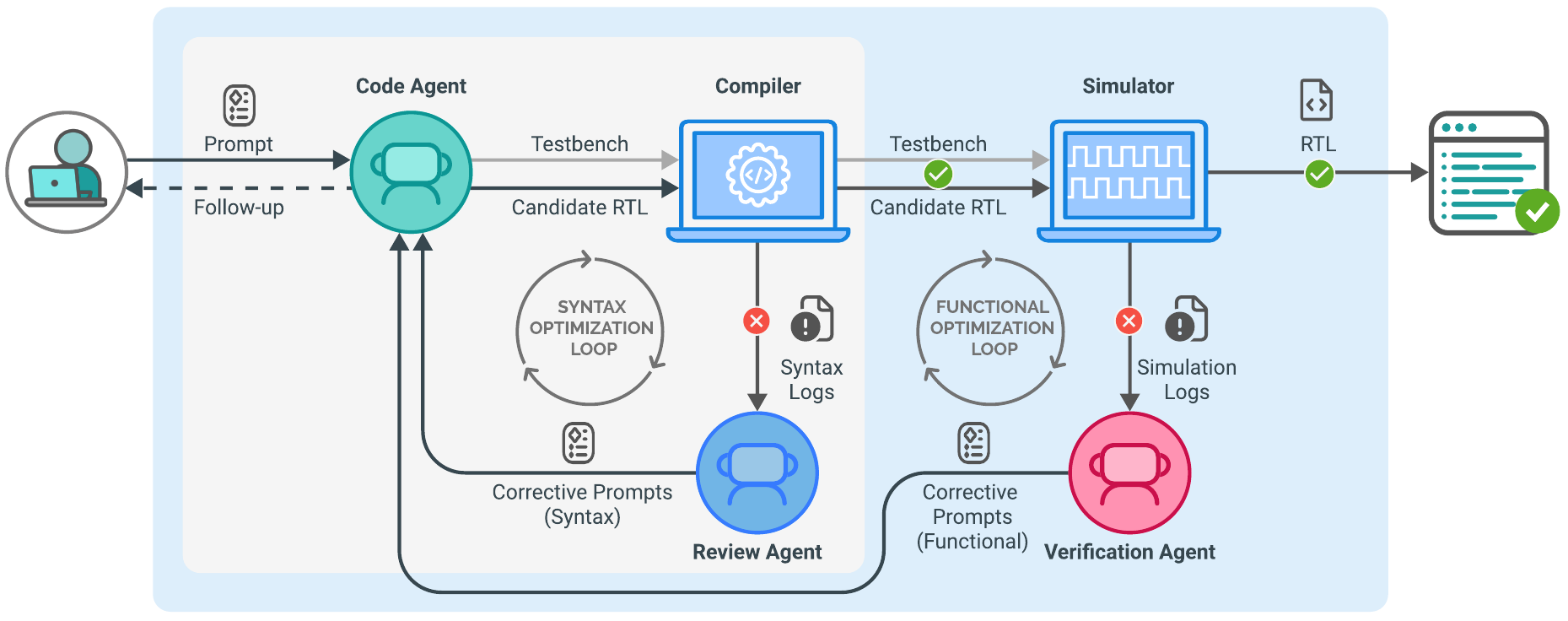}
    \caption{Architecture overview of the proposed {\scshape AIvril}2 framework.}
    \label{fig:overall_flow}
\end{figure*}

In this section, we detail the internal mechanisms of the proposed framework: a two-stage LLM-aware RTL design methodology with enforced functional verification. The overall structure is depicted in Fig.~\ref{fig:overall_flow}. The framework is built around two key loops: the {\em Syntax Optimization} and the {\em Functional Optimization} loop, each governed by three agents: the {\em Code Agent}, {\em Review Agent}, and {\em Verification Agent}. The {\em Syntax Optimization} loop is supervised by the Review Agent, while the {\em Functional Optimization} one falls under the purview of the {\em Verification Agent}, as detailed in the following.
\subsection{Code Agent}
\begin{figure*}[t]
    \centering
    \includegraphics[width=\textwidth]{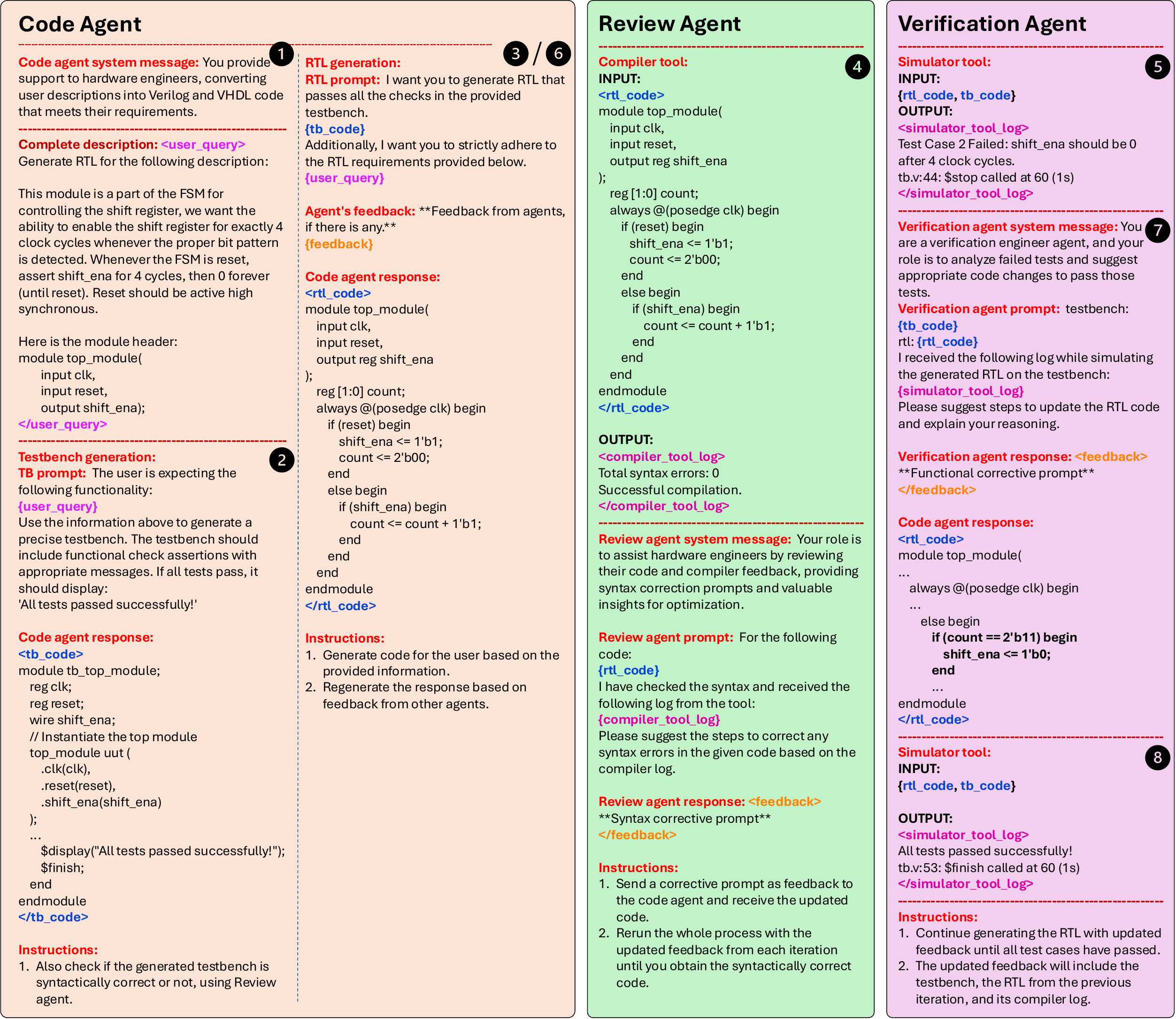}
    \caption{Practical example of the proposed workflow and internal state representation of the agents in {\scshape AIvril}2.}
    \label{fig:overall_flow_example}
\end{figure*}

The {\em Code Agent} acts as the primary code generation component within {\scshape AIvril}2, and it is responsible for translating user requirements into functional RTL code. As the only source of code generation throughout the process, it ensures design consistency and coherence. The agent begins by analyzing the user-provided prompt, which outlines the desired RTL design functionality. For reference, Figure~\ref{fig:overall_flow_example} illustrates an example involving requirements for a shift register (box denoted with \ballnumber{1}). If the prompt lacks sufficient detail, the {\em Code Agent} initiates an interactive dialogue with the user to gather further information. In practice, this behavior is accomplished via additional {\em ad hoc} system prompts. Once a complete prompt is provided, the agent first generates a comprehensive testbench based on the received specifications, ensuring that all potential test cases, that a functionally correct RTL design must pass, are covered. This represents our self-verification approach, an example of which is illustrated in Figure~\ref{fig:overall_flow_example}, step \ballnumber{2}. This step is critical, as it sets the baseline for the subsequent verification process. Using both the user prompt and the generated testbench as references, the {\em Code Agent} then produces an initial version of the RTL code (step \ballnumber{3}). As shown in the example, this initial code is forwarded to the {\em Review Agent} for syntax checking and further validation. At each iteration, the {\em Code Agent} plays a key role in refining the RTL design. It incorporates feedback from other agents in the form of corrective prompts, implicitly managing different versions of the RTL code throughout the iterative process, thereby facilitating easy change tracking and also enabling rollbacks when necessary.

\subsection{Review Agent}
The main role of the {\em Review Agent} is to ensure the syntactical correctness of the generated code. Capable of integrating with any industry-standard RTL compiler, its primary function is to meticulously review the code for syntax errors and provide detailed feedback, distilled from compilation logs, to the {\em Code Agent}. As illustrated in Figure~\ref{fig:overall_flow_example}, the {\em Review Agent} receives the initial code produced by the {\em Code Agent} based on the user's design specifications. Upon compilation, a comprehensive log file is generated, which serves as the primary input for the {\em Review Agent}'s analysis. At this stage, the {\em Review Agent} performs an in-depth examination of the output. It analyzes the logs for any signs of syntax errors, carefully parsing the information to identify specific issues within the code. This analysis goes beyond simple error detection: the agent also identifies the exact locations of errors by extracting line numbers and relevant code snippets from the log file, with the ultimate goal of converting them into corrective prompts. In the example shown in Figure~\ref{fig:overall_flow_example}, the {\em Review Agent}'s analysis finds no syntax errors in the initial code (step \ballnumber{4}). This successful syntax check allows the process to move directly to the functional verification stage. On the other hand, if syntax errors are detected, the agent generates a highly detailed and actionable corrective prompt. This prompt provides a comprehensive breakdown of each syntax error, including the exact line numbers where the errors occur, relevant code snippets surrounding the problematic areas, and potential suggestions or hints for resolving the syntax issues. It is important to note that the level of detail provided is crucial, as it allows the {\em Code Agent} to quickly identify and correct syntax issues in the minimum number of iterations. The {\em Review Agent} continues this process iteratively, working in close coordination with the {\em Code Agent} until the code achieves syntactical correctness. This iterative refinement ensures that the code progresses to the functional verification stage only after it has been thoroughly validated for syntax errors.

\subsection{Verification Agent}
The {\em Verification Agent} represents the final stage in our framework and is responsible for ensuring the functional correctness of the RTL design. This agent triggers once both the RTL code and the testbench have been validated as syntactically correct by the {\em Review Agent}. The primary objective of the {\em Verification Agent} is to verify that the generated RTL code passes all the test cases outlined in the testbench, as generated at step \ballnumber{3} or \ballnumber{6}, depending on the stage of the process. As also shown in Figure~\ref{fig:overall_flow_example}, the {\em Verification Agent}'s workflow begins with the simulation process. The agent then analyzes the resulting simulation logs in order to detect any discrepancies between the expected and the actual outputs. In the given example, the initial simulation identifies a functional error: ``Test Case 2 Failed: shift\_ena should be 0 after 4 clock cycles,'' as detailed in step \ballnumber{5}. Based on this analysis, the {\em Verification Agent} generates a corrective prompt to guide the {\em Code Agent} in resolving the functional issues identified during the simulation (steps \ballnumber{6} and \ballnumber{7}). A critical aspect of this verification workflow is the consistent use of the same testbench throughout all iterations. While the RTL code may undergo multiple revisions based on feedback, the testbench remains unchanged. This approach ensures a standardized and unbiased evaluation of each RTL version, allowing for precise tracking of improvements and consistent verification in meeting the original functional requirements. The {\em Verification Agent} operates in synergy with the {\em Code Agent}, providing feedback and receiving updated RTL designs until either all test cases pass successfully or a predefined maximum number of iterations is reached. In the example, this iterative process is highlighted when, after re-verification following {\em Code Agent}'s refinements, the output log confirms that ``All tests passed successfully!'' (step \ballnumber{8}). This outcome indicates that the RTL now fully satisfies the user’s requirements, demonstrating the efficiency of the feedback loop between the {\em Verification} and the {\em Code Agent}.

\section{Experimental Results}\label{sec:results}

In this section, we present the experimental evaluation of the proposed {\scshape AIvril}2 framework. Our goal is to rigorously assess the performance and robustness of the tool across a diverse set of benchmarks, ensuring a thorough and unbiased analysis of its capabilities. To achieve this, we selected key evaluation metrics that emphasize the strengths of our approach in realistic scenarios. Specifically, we focused on metrics that address both syntactical and functional correctness, providing a comprehensive perspective on the effectiveness of our solution in handling complex design tasks.

\subsection{Methodology}
We employed all 156 benchmarks from the VerilogEval-Human benchmark suite~\cite{liu2023verilogeval} in our experiments, which enabled us to encompass a broad range of design complexities. For both optimization loops, performance was evaluated using the unbiased pass@$k$ estimator (with $k=1$), as described in~\cite{chen2021evaluating}. We distinguish between pass@$k_{\mathcal{S}}$, which represents the success rate of designs passing all syntax checks, and pass@$k_{\mathcal{F}}$, which reflects the success rate of designs that are not only syntactically correct but also functionally accurate. Notably, pass@$k_{\mathcal{F}}$ was determined by executing the testbenches provided in the benchmark suite, ensuring a comprehensive validation of the overall approach.

For both the syntax check and functional simulation stages, we utilized Vivado Design Suite - HLx Editions 2018.1, as to easily enable mixed-language simulations. To gain broader insights into the capabilities of various LLMs in generating RTL, we employed different models for the agents: Claude 3.5 Sonnet~\cite{claude3.5}, GPT-4o~\cite{gpt4o}, and Llama3-70B~\cite{dubey2024llama}. All models were used without any fine-tuning or RAG integration. Additionally, the {\em temperature} and {\em top\_p} parameters for each LLM were set to 0.2 and 0.1, respectively.

\subsection{Results \& Discussion}
\begin{table*}[!h]
    \centering
    \resizebox{1\textwidth}{!}{
        \normalsize
        \addtolength\tabcolsep{8pt}
    \begin{tabular}{lrrrrrr}\toprule
    \multirow{2.5}{*}{\textbf{Technology}} &\multicolumn{3}{c}{\textbf{Verilog}} &\multicolumn{3}{c}{\textbf{VHDL}} \\\cmidrule(lr){2-4}\cmidrule(lr){5-7}
    &\textbf{pass@$1_{\mathcal{S}}$} &\textbf{pass@$1_{\mathcal{F}}$} &\textbf{$\Delta_{\mathcal{F}}$} &\textbf{pass@$1_{\mathcal{S}}$} &\textbf{pass@$1_{\mathcal{F}}$} &\textbf{$\Delta_{\mathcal{F}}$} \\\midrule
    Llama3-70B &71.15 &37.82 &- &1.28 &0 &- \\
    GPT-4o &71.79 &51.29 &- &39.1 &27.56 &- \\
    Claude 3.5 Sonnet &91.03 &60.23 &- &88.46 &53.85 &- \\\midrule
    \textbf{{\scshape AIvril}2 (Llama3-70B)} &100 &55.13 &45.76 &58.87 &32.69 &N/A \\
    \textbf{{\scshape AIvril}2 (GPT-4o)} &100 &72.44 &41.23 &100 &59.62 &116.32 \\
    \textbf{{\scshape AIvril}2 (Claude 3.5 Sonnet)} &100 &77 &27.84 &100 &66 &22.56 \\
    \bottomrule
    \textbf{Average} & & &\textbf{38.28} &\multicolumn{2}{c}{} &\textbf{$\gg$ 69.44} \\
    \bottomrule
    \end{tabular}
    }
    \caption{Summary of pass-rate results, with column {\bf $\Delta_{\mathcal{F}}$} showing the percentage improvement of the proposed technique over the corresponding baseline model in terms of functional pass rate. All values are expressed as percentages.}
    \label{tab:results}
\end{table*}
Table~\ref{tab:results} reports the experimental results. The table is organized as follows: for each of the two target RTL languages, we report the syntax pass rate (column pass@$1_{\mathcal{S}}$), the functional pass rate (column pass@$1_{\mathcal{F}}$), and the improvements, in terms of functional pass rate, of {\scshape AIvril}2 w.r.t. the corresponding baseline LLM (column {\bf $\Delta_{\mathcal{F}}$}). As the data suggest, in all configurations, our framework achieves a pass@$1_{\mathcal{S}}$ of 100\%, except in one case: Llama3-70B for VHDL, which attained a pass@$1_{\mathcal{S}}$ of only 58.87\%. While not perfect, this result still demonstrates the efficacy of the {\em Syntax Optimization} loop, considering that the baseline model only achieved a pass@$1_{\mathcal{S}}$ of 1.28\%, marking an impressive 46$\times$ improvement in code quality. This outcome highlights a significant gap in Llama3-70B's foundation knowledge concerning VHDL design, likely due to the scarcity of VHDL source code included in its training data. Consequently, the baseline pass@$1_{\mathcal{F}}$ for Llama3-70B is 0\%. However, even in this case, {\scshape AIvril}2 was able to recover functional correctness in the analyzed VHDL designs, achieving a pass@$1_{\mathcal{F}}$ of 32.69\%. More broadly, our approach improved RTL code quality generated by the baseline LLMs by 38.28\% for Verilog and at least 69.44\% for VHDL, on average. These results highlight the effectiveness of the proposed framework in enhancing both syntactical and functional correctness of RTL designs across all LLMs and RTL languages. The significant improvements in VHDL, despite the models' initial disadvantages, underscore the capability of {\scshape AIvril}2 to boost performance across different hardware description languages, even when the training data for the underlying model is imbalanced.

\begin{figure}[h]
    \centering
    \includegraphics[width=.8\linewidth]{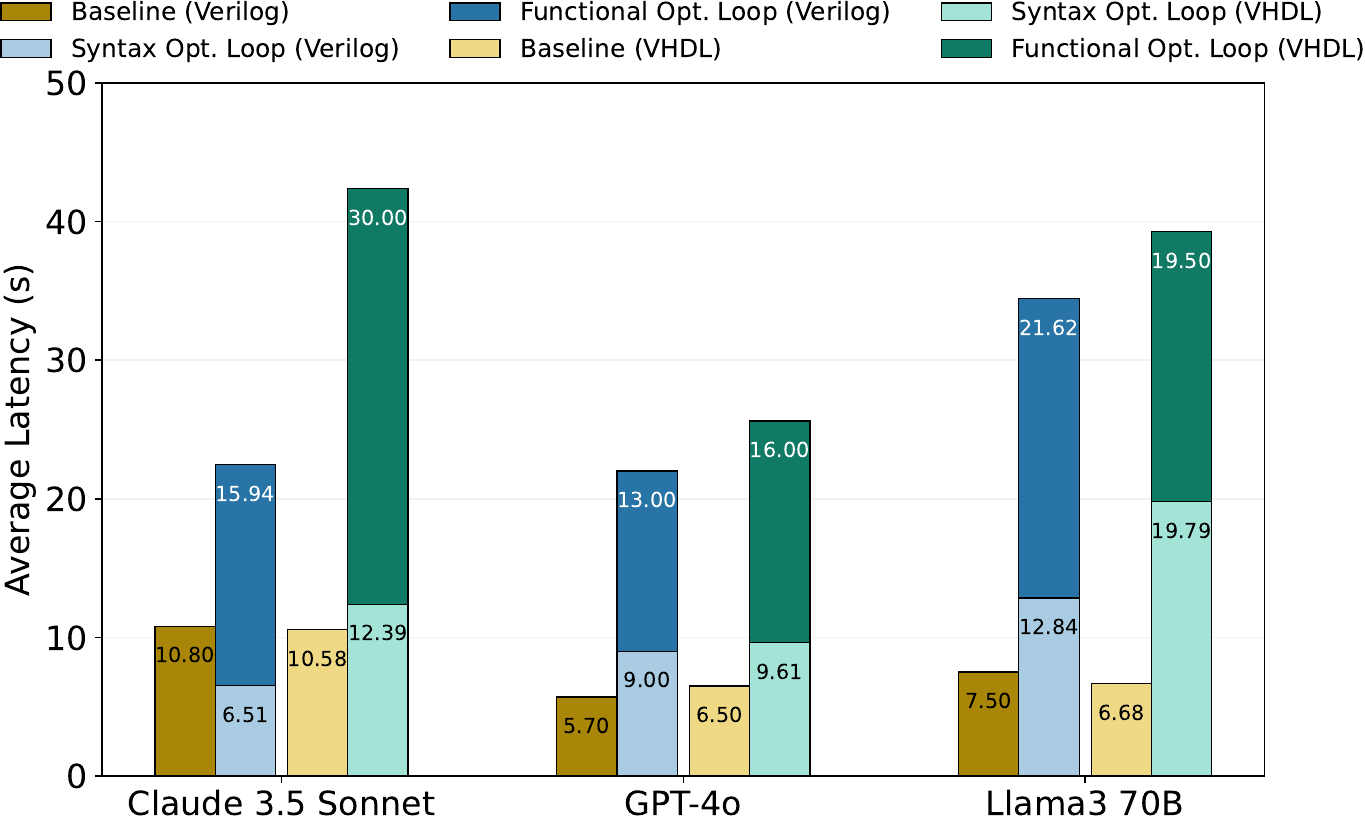}
    \caption{Average latency breakdown across optimization loops for the proposed framework. Reported figures account for the execution times of EDA tools.}
    \label{fig:avg_exec_time}
\end{figure}

Architectural differences among LLMs often lead to varying execution times. This is particularly important to consider when applying optimization loops around LLMs, as it helps validate the practicality of our approach in real-world scenarios. Figure~\ref{fig:avg_exec_time} illustrates the average latency for different LLMs, as well as a breakdown of the execution times for both optimization loops across all considered LLM configurations. The most significant latency increase was observed with Llama3-70B when generating VHDL. As the plot suggests, the latency gap from the baseline in this case is approximately 6$\times$ (e.g., 6.68 {\em vs.} 39.29 seconds). This is partly due to the higher number of iterations required by Llama3-70B to converge towards a solution. More specifically, Llama3-70B required an average of 3.95 cycles for the {\em Syntax Optimization} loop and 4.7 cycles for the {\em Functional Optimization} loop to converge. In contrast, the smallest latency increase was recorded with Claude 3.5 Sonnet for Verilog generation, which showed roughly a 2$\times$ increase in execution time. This configuration required an average of 2 steps for the {\em Syntax Optimization} loop and 3 steps for the {\em Functional Optimization} loop. Overall, while some latency gaps may appear significant, it is worth emphasizing that the worst-case average latency introduced by our approach did not exceed 42 seconds. This is a reasonable trade-off, considering the substantial time saved in avoiding potential manual debugging and verification. Another noteworthy observation is the latency recorded for Claude 3.5 Sonnet during the {\em Functional Optimization} loop for VHDL, which was the highest among all solutions. This increased latency can be attributed to the LLM's own processing, particularly due to the higher complexity induced by corrective prompts.

\subsection{Comparison with State-of-the-Art Approaches}\label{sec:sota-comparison}
As already discussed in Section~\ref{sec:related-work}, recent frameworks and fine-tuned LLMs have been introduced to enhance RTL code quality within the context of GenAI solutions. Table~\ref{tab:sota-comparison} provides a comparison between our proposed framework and existing solutions. For each technique, we report the license associated with the adopted LLM and the pass@$1_{\mathcal{F}}$ metric. Due to limited data availability, our comparison focuses on Verilog generation only, as, to the best of the authors' knowledge, this work is the first to evaluate GenAI solutions for VHDL. As shown in the table, our approach outperforms existing solutions in both open-source and closed-source regimes. Most notably, the highest performance gap is recorded w.r.t. ChipNemo-13B~\cite{liu2023chipnemo}, where our solution achieves a ~3.4$\times$ higher pass@$1_{\mathcal{F}}$. These results further highlight the strength of {\scshape AIvril}2 in achieving state-of-the-art performance in RTL generation.

\begin{table*}[!h]
    \centering
    \resizebox{.6\textwidth}{!}{
        \normalsize
        \addtolength\tabcolsep{2pt}
        \begin{tabular}{lrrrr}\toprule
            \textbf{Technology} &\textbf{Model License} &\textbf{pass@$1_{\mathcal{F}}$ (\%)} \\\midrule
            Llama3-70B~\cite{dubey2024llama} &\multirow{3}{*}{Open Source} &37.82 \\
            CodeGen-16B~\cite{thakur2023benchmarking} & &41.9 \\
            CodeV-CodeQwen~\cite{zhao2024codev} & &53.2 \\ \midrule
            ChipNemo-13B~\cite{liu2023chipnemo}&\multirow{8}{*}{Closed Source} &22.4 \\
            ChipNemo-70B~\cite{liu2023chipnemo} & &27.6 \\
            CodeGen-16B-Verilog-SFT~\cite{liu2023verilogeval} & &28.8 \\
            RTLFixer~\cite{tsai2023rtlfixer} & &36.8\\
            VeriAssist~\cite{huang2024towards} & &50.5\\
            GPT-4o~\cite{gpt4o} & &51.29 \\
            Claude 3.5 Sonnet~\cite{claude3.5} & &60.23 \\
            {\scshape AIvril}~\cite{sami2024aivril} & &67.3\\\midrule\midrule
            \textbf{{\scshape AIvril}2 (Llama3-70B)} &Open Source &\textbf{55.13} \\\midrule
            \textbf{{\scshape AIvril}2 (GPT-4o)} &\multirow{2}{*}{Closed Source} &\textbf{72.44} \\
            \textbf{{\scshape AIvril}2 (Claude 3.5 Sonnet)} & &\textbf{77} \\
            \bottomrule
            \end{tabular}
    }
    \caption{Comparison of state-of-the-art RTL generation techniques. Column pass@$1_{\mathcal{F}}$ only reports numbers for Verilog.}
    \label{tab:sota-comparison}
\end{table*}

\section{Conclusions}\label{sec:conclusions}
In this work, we introduced {\scshape AIvril}2, a novel self-verifying, LLM-aware RTL design framework that entails a {\em Syntax Optimization} and a {\em Functional Verification} loop to enhance syntax and functional correctness. Experimental results demonstrated that our framework significantly improves code quality across a wide range of benchmarks, outperforming baseline models and prior solutions in both Verilog and VHDL generation. Despite some added latency, the overall execution times remained reasonable, making the trade-off worthwhile given the reduction in manual verification. These results highlight the robustness and versatility of the proposed framework, paving the way for future advancements in automated GenAI for RTL.

\bibliographystyle{IEEEtran}
\bibliography{refs}
\end{spacing}

\end{document}